\documentclass[%
superscriptaddress,
twocolumn,
 amsmath,amssymb,
 aps,
prb,
]{revtex4-1}

\usepackage{graphicx}
\usepackage{bm}
\usepackage{hyperref}

\renewcommand{\v}[1]{\ensuremath{\mathbf{#1}}} 

\newcommand{\abs}[1]{\left| #1 \right|} 
\newcommand{\pd}[2]{\frac{\partial #1}{\partial #2}}

\let\baraccent=\= 
\renewcommand{\=}[1]{\stackrel{#1}{=}} 

\begin{document}

\title{Interplay of the orbital magnetic moment and the chiral magnetic effect in Shubnikov-de Haas quantum oscillations}

\author{Jinho Yang}

\affiliation{Department of Physics, POSTECH, Pohang, Gyeongbuk 37673, Korea}

\author{Ki-seok Kim}

\affiliation{Department of Physics, POSTECH, Pohang, Gyeongbuk 37673, Korea}

\affiliation{Asia Pacific Center for Theoretical Physics (APCTP), Pohang, Gyeongbuk 37673, Korea}

\date{\today}

\begin{abstract}
Emergent Lorentz symmetry and chiral anomaly are well known to play an essential role in anomalous transport phenomena of Weyl metals. In particular, the former causes a Berry-curvature induced orbital magnetic moment to modify the group velocity of Weyl electrons, and the latter results in the chiral magnetic effect to be responsible for a ``dissipationless" longitudinal current channel of the bulk. In this study, we verify that intertwined these two effects can be measured in Shubnikov-de Haas (SdH) quantum oscillations, where a double-peak structure of the SdH oscillation appears to cause a kink in the Landau fan diagram. We examine three different cases which cover all possible experimental situations of external electric/magnetic fields and identify the experimental condition for the existence of the double-peak structure. We claim that interplay of the orbital magnetic moment and the chiral magnetic effect in SdH quantum oscillations is an interesting feature of the Weyl metal state.
\end{abstract}

\maketitle

\section{Introduction}

``Landau level" further splitting in Shubnikov-de Haas (SdH) quantum oscillations is commonly observed in topological materials such as Dirac metals \cite{Cd3As2Jeon, Cd3As2Cao, ZrTe5Chen, ZrTe5Liu, ZrSiSHu, ZrTe5Zheng} and Weyl metals \cite{NbAsYuan}. This double-peak structure in the SdH oscillation gives rise to a kink signature in the Landau fan diagram. This study suggests the origin of the Landau-level splitting in a Weyl metallic state.

It is well established that emergent Lorentz symmetry and chiral anomaly \cite{WM1,WM2,WM3,WM4} play a central role in anomalous transport phenomena of Weyl metals \cite{WM_Review1,WM_Review2,WM_Review3}. The relativistic invariance enforces that the total angular momentum given by sum of the spin and orbital angular momentum has to be conserved. In other words, the Lorentz boost changes not only the spin angular momentum but also the orbital one. As a result, a Weyl electron away from the Weyl point carries an effective angular momentum proportional to the Berry curvature at each momentum position. This Berry-curvature induced orbital angular momentum causes an additional energy contribution given by an effective Zeeman coupling form between the Berry curvature and an external magnetic field \cite{gv1,gv2,Moore}. This effective Zeeman energy changes the group velocity depending on the chirality, which is reduced (enhanced) in the positive (negative) chirality Weyl point.

The chiral anomaly means that U(1) chiral currents given by positive-chirality Weyl-electron currents minus negative ones cannot be preserved within the quantum mechanical principle as long as U(1) charge currents are enforced to be conserved \cite{Anomalies in QFT,TRB WM Anomaly,Iksu_Kiseok_Anomaly}. Physical realization of the chiral anomaly is that there exists a ``dissipationless" current channel in the bulk, which gives rise to charge pumping from a positive-chiral Fermi surface to a negative-chiral one. More precisely, the time evolution of the chiral charge is given by the chemical potential difference between the positive and negative chiral Fermi surfaces, referred to as the chiral chemical potential $\mu_{5} \propto \bm{E} \cdot \bm{B}$ \cite{CME_I,CME_II,CME_III,CME_IV,CME_V,CME_VI,CME_VII,CME_VIII,CME_IX,CME_X,CME_XII,CME_XIII,CME_XIV,CME_XV,CME_XVI,mu5}, where $\bm{E}$ and $\bm{B}$ are externally applied electric and magnetic fields, respectively. This so called chiral magnetic effect is responsible for $|\bm{B}|^{2}$ enhancement of longitudinal magnetoconductivity \cite{TSB_WM1,TSB_WM2,ISB_WM1,ISB_WM2,ISB_WM3,ISB_WM4,ISB_WM5,ISB_WM6,ISB_WM7}.

%
%

In this study, we demonstrate that these two effects are intertwined to cause a double-peak structure in the SdH quantum oscillation. In particular, we find criteria on the existence of this double-peak structure in a time-reversal symmetry-broken Weyl metal state. This leads us to manipulate the splitting structure as a function of the external magnetic field in the linear-response regime. We claim that the interplay of the orbital magnetic moment and the chiral magnetic effect in SdH quantum oscillations is an interesting feature of the Weyl metal state.

\section{Electrical conductivity of a time-reversal symmetry-broken Weyl metal state with a pair of chiral Fermi surfaces}

\subsection{General formula of conductivity for SdH quantum oscillations}

Transverse ($\v{E} \perp \v{B}$) and longitudinal ($\v{E} \parallel \v{B}$)  quantum oscillations of metals under external magnetic fields are given by
\begin{widetext}
\begin{eqnarray}
\sigma_{xx} &=& \sigma_{xx}^{l=0} + 2 \sum_{l = 1}^{\infty} e^{-\lambda_D l} \sigma_{xx}^{(l)}\left(\cos{\left(\frac{\pi l}{2\zeta_F}+\frac{\pi}{4} - l\phi\right)}+\frac{l\pi}{\zeta_F}\cos{\left(\frac{\pi l}{2\zeta_F}-\frac{\pi}{4} -l\phi \right)}\right) \label{sigmaxx} \\
\sigma_{zz} &=& \sigma_{zz}^{l=0} + 2 \sum_{l = 1}^{\infty} e^{-\lambda_D l} \sigma_{zz}^{(l)}\left(\cos{\left(\frac{\pi l}{2\zeta_F}+\frac{\pi}{4} -l\phi \right)}\right) \label{sigmazz}
\end{eqnarray}
\end{widetext}
in the semi-classical limit \cite{Lifshits, sigmaz}. Here, $\sigma^{l=0}_{xx (zz)}$ is a non-oscillatory conductivity term as a function of the external magnetic field. $\sigma^{l \not= 0}_{xx (zz)}$ is an amplitude of the SdH quantum oscillation, where $l$ is an integer. The complete form of $\sigma^{l \not= 0}_{xx (zz)}$ is shown in appendix and below sections. $\lambda_D = \frac{\pi \hbar/\tau_q}{\hbar k_F v_F \zeta_F}$ is the Dingle damping factor, where $k_F$, $v_F$, and $\tau_q$ are Fermi momentum, Fermi velocity, and relaxation time (given by forward scattering mostly) of electrons, respectively. $\zeta_F =  \frac{eB}{2 \hbar k_F^2}$ is a dimensionless length scale given by the magnetic length and the Fermi momentum. $\phi = 2 \pi \gamma$ is a phase shift from the Sommerfeld-Bohr quantization condition. $\gamma$ is $1/2$ in a conventional metal whereas it is $0$ in the presence of the Berry phase $\Phi_B = \pi$, more precisely, given by $\gamma = \frac{1}{2} - \frac{\Phi_B}{2 \pi}$. At present, we do not consider $l=0$ terms, and focus on oscillatory components as a function of the external magnetic field.


\subsection{Key feature of a Weyl metal phase for SdH quantum oscillations: Chirality-dependent Fermi-momentum change}

An essential point in the SdH quantum oscillation of the Weyl metal phase is that there are two Fermi momenta depending on the chirality, which originates from two reasons: i) Berry-curvature induced orbital-magnetic-moment gives rise to an additional Zeeman energy contribution under the external magnetic field, modifying the group velocity in a chirality-dependent way \cite{gv1,gv2,Moore}, and ii) the chiral chemical potential $\mu_5$ appears to realize the chiral anomaly, referred to as the chiral magnetic effect \cite{CME_I,CME_II,CME_III,CME_IV,CME_V,CME_VI,CME_VII,CME_VIII,CME_IX,CME_X,CME_XII,CME_XIII,CME_XIV,CME_XV,CME_XVI,mu5}.

\begin{figure}
\centering
\includegraphics[width=9cm]{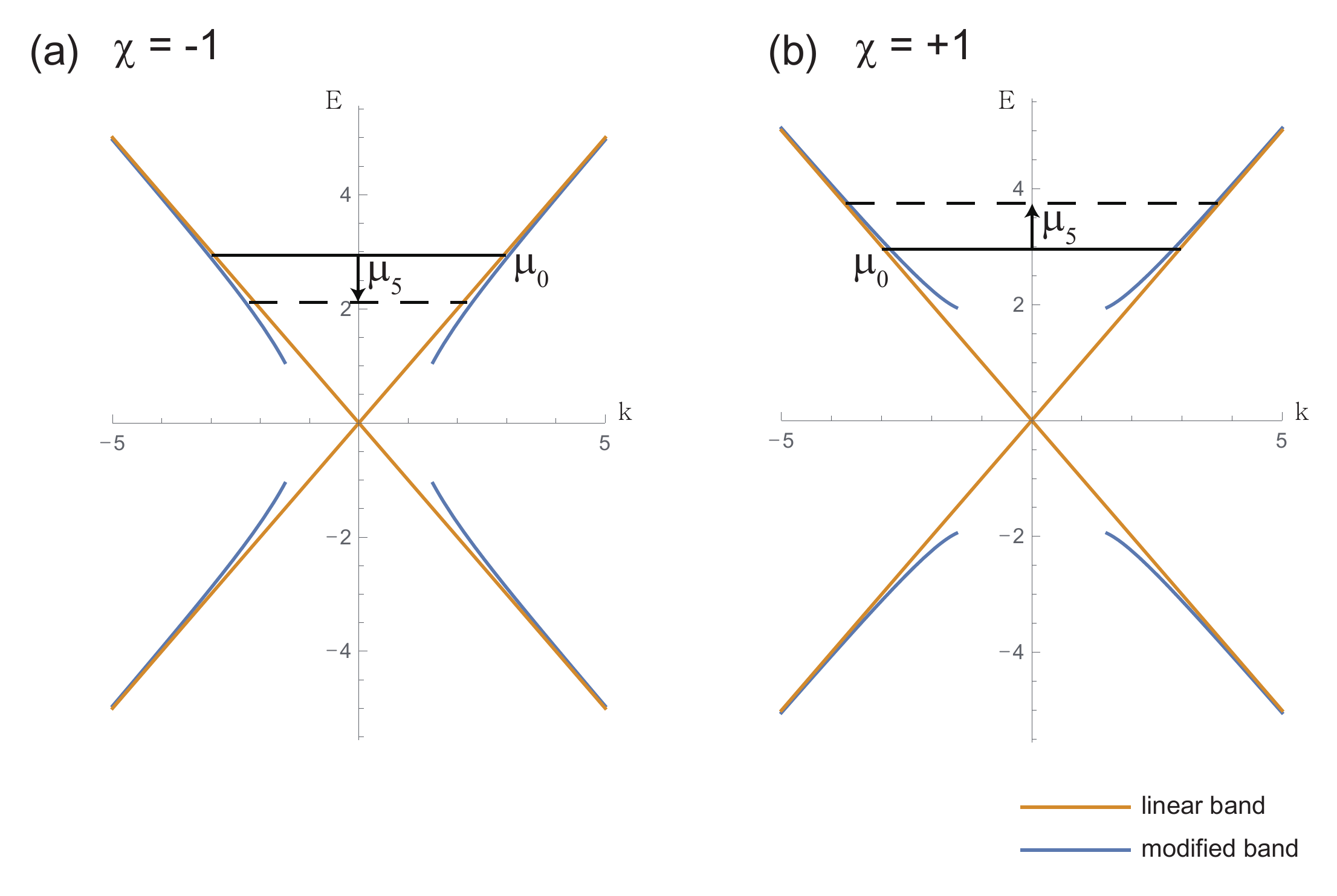}
\caption{Dispersion relations for a pair of chiral Fermi surfaces. They are modified by the Zeeman-energy contribution from the Berry-curvature induced orbital magnetic moment. We also point out that the area of each chiral Fermi surface is further different, which results from the chiral chemical potential $\mu_5$.}
\end{figure}

%
%

Although the linear band structure of $\epsilon(\v{k}) = \hbar v_F |\v{k}|$ is taken into account in Weyl materials, it turns out that this expression is not complete in the respect of the Lorentz invariance. An important point is that the spin angular momentum is assigned to each momentum point of the chiral Fermi surface by spin-momentum locking or spin enslavement. If one considers the Lorentz boost, the spin angular momentum has to be changed. On the other hand, the total angular momentum is conserved. In this respect there must be an orbital angular momentum to compensate the change of the spin angular momentum. Actually, the Berry curvature turns out to play the role of the orbital angular momentum. This emergent orbital angular momentum gives rise to an additional Zeeman energy contribution. As a result, the dispersion relation is modified as \cite{gv1,gv2,Moore}
\begin{eqnarray}
\epsilon(\v{k}) = \hbar \big(v_F - \frac{e}{\hbar}(\v{\Omega} \cdot \v{v_F}) B \big) |\v{k}| ,
\end{eqnarray}
where $\boldsymbol{\Omega} = \chi \frac{\hat{\v{k}}}{2 k^2}$ is the Berry curvature for the chirality $\chi = \pm 1$.

If there is no external magnetic field ($\v{B}$), the Fermi energy is given by
\begin{eqnarray}
\mu_{0} &=& \hbar v_F k_{F0}.
\end{eqnarray}
When there exists an external magnetic field, the Fermi momentum ($k_F^\pm$) for each chiral Fermi surface ($\chi = \pm 1$) has to be modified as
\begin{eqnarray}
\mu_{0} &=& \hbar v_F k_{F0} =  \hbar k_F^\pm \big(v_F - \frac{e}{\hbar}(\v{\Omega_{\pm}} \cdot \v{v_F}) B \big). \label{changeddispersion}
\end{eqnarray}
Now, we get the Fermi momentum for each chirality with the external magnetic field in the following way\begin{eqnarray}
k_F^\pm &=& \frac{k_{F0}}{2} (1+ \sqrt{1\pm \frac{2eB}{\hbar k_{F0}^2}}) \nonumber \\
&=& \frac{k_{F0}}{2} (1+\sqrt{1\pm 4\zeta_{F}}) \nonumber \\
&\approx& k_{F0} (1 \pm \zeta_F) . \label{zetaFcorrection}
\end{eqnarray}
Here, $k_{F0}$ is the Fermi momentum without an external magnetic field (the same Fermi momentum for each chirality in this case) whereas $k_F^\pm$ is the chirality-dependent Fermi momentum with an external magnetic field. We get the $\pm \zeta_F$ correction due to the dispersion change.

%
%

Now, we are going to turn on the $\v{E}$ field. As discussed before, the chiral anomaly is realized by chiral charge pumping through a dissipationless current channel when both $\v{E}$ and $\v{B}$ fields are applied simultaneously in the parallel direction. This chiral magnetic effect is given by the chiral chemical potential \cite{CME_I,CME_II,CME_III,CME_IV,CME_V,CME_VI,CME_VII,CME_VIII,CME_IX,CME_X,CME_XII,CME_XIII,CME_XIV,CME_XV,CME_XVI,mu5}
\begin{eqnarray}
\mu_5 &=& \frac{\mu_+ - \mu_-}{2} \nonumber
\\ &=& \frac{3}{4} \frac{v_F^3}{\pi^2}\frac{e^2}{\hbar^2 c} \Big(\frac{\v{E} \cdot \v{B}}{T^2+\mu_0^2/\pi^2} \Big) \tau_v  \nonumber \\
&\equiv& \hbar v_f a \v{E} \cdot \v{B} \ll \mu_0 , \label{mu5eqn}
\end{eqnarray}
where $\tau_v$ is inter-valley scattering time. As a result, the chemical potential for each chiral Fermi surface is given by
\begin{eqnarray}
\mu_\pm &=&  \mu_0 \pm \mu_5 \nonumber \\
&=& \hbar v_F k_{F0}  \pm \hbar v_F a \v{E} \cdot \v{B} \nonumber \\
&=& \hbar k_{\mu5}^\pm \Big(v_f \mp \frac{e B v_f}{2 (k_{\mu5}^\pm)^2 \hbar} \Big) . \label{getkmu5eqn}
\end{eqnarray}
Here, we assumed that the dispersion of each chirality is maintained as Eq. (\ref{changeddispersion}) even though their chemical potentials are modified from $\mu_0$.

Solving Eq. (\ref{getkmu5eqn}), we obtain further modifications of the Fermi momentum $k^{\pm}_{\mu5}$ due to $\mu_5$ as
\begin{eqnarray}
k_{\mu5}^\pm &\approx& k_{F0} (1 \pm \frac{eB}{2 \hbar k_{F0}^2} \pm \frac{a EB}{k_{F0}}) = k_{F0} (1 \pm \zeta_F \pm \frac{\mu_5}{\mu_0}) \nonumber \\
&=& k_{F0}\pm \delta k . \label{mu5correction}
\end{eqnarray}
We recall that the $\mu_5/\mu_0$ correction in $\delta k = k_{F0}(\zeta_F + \frac{\mu_5}{\mu_0})$ comes from the chemical potential change, whereas the $\zeta_F$ correction results from the band dispersion change.

\subsection{SdH quantum oscillations from two types of chiral Fermi surfaces}

Introducing the chirality-dependent Fermi momentum $k_{\mu5}^{\pm}$ into the longitudinal conductivity Eq. (\ref{sigmazz}), we obtain the SdH quantum oscillation for each chiral Fermi surface as follows
\begin{widetext}
\begin{eqnarray}
\sigma_{zz}^{osc \pm} &=& 2 \sum_l e^{- l \lambda_D^\pm} \sigma_{zz}^{(l)\pm}\left(\cos{\left(\frac{\pi l}{2\zeta_F^\pm}+\frac{\pi}{4}\right)}\right) \nonumber \\
&=& \sum_l \frac{C_1 B^{\frac{1}{2}}}{\sqrt{l}\sinh(l \frac{C_3 k_{F0}}{B}(1 \pm \delta k/k_{F0}))} e^{-l \frac{C_2 k_{F0}}{B} (1 \pm \delta k/k_{F0})} \cos\left(\frac{l\pi \hbar (k_{\mu5}^\pm)^2}{eB} + \frac{\pi}{4} \right) \nonumber \\
&=& \sum_l \frac{C_1 B^{\frac{1}{2}}}{\sqrt{l}\sinh(l \frac{C_3 k_{F0}}{B}(1 \pm \delta k/k_{F0}))} e^{-l \frac{C_2 k_{F0}}{B} (1 \pm \delta k/k_{F0})} \cos\left(\frac{l\pi \hbar}{eB}(k_{F0}^2 \pm 2 k_{F0} \delta k + \delta k^2) + \frac{\pi}{4} \right) \nonumber \\
&=& \sum_l \frac{C_1 B^{\frac{1}{2}}}{\sqrt{l}\sinh(l \lambda(1 \pm \delta k/k_{F0}))} e^{-l \lambda_D (1 \pm \delta k/k_{F0})} \cos\left(\frac{l\pi \hbar k_{F0}^2}{eB}+ l\Delta_\pm (E,B) + \frac{\pi}{4} \right) ,
\end{eqnarray}
where $C_1 = \frac{\tau e k_B T}{\hbar \pi}\left(\frac{e}{\hbar}\right)^\frac{3}{2}, \quad C_2 = \frac{2 \pi \hbar}{\tau_q v_f e}, \quad C_3 =  \frac{2 \pi^2 k_B T}{v_f e}, \quad \lambda_D = \frac{C_2 k_{F0}}{B} = \frac{\pi \hbar/\tau_q}{\hbar k_F v_F \zeta_F}, \quad \lambda = \frac{C_3 k_{F0}}{B} = \frac{\pi^2 T}{\hbar k_F v_F \zeta_F}$. Here, we observe a key control parameter $\Delta_\pm (E,B)$ in terms of $\delta k = k_{F0}(\zeta_F + \frac{\mu_5}{\mu_0})$, more explicitly given by
\begin{eqnarray}
\Delta_\pm (E,B) &=& 2\frac{\pi \hbar k_{F0}^2}{eB}\left \{ \pm (a\v{E}\cdot \v{B}/k_{F0} + \frac{eB}{2\hbar k_{F0}^2})+\frac{1}{2}(a\v{E}\cdot \v{B}/k_{F0} + \frac{eB}{2\hbar k_{F0}^2})^2\right \} \nonumber \\
&=& \frac{\pi}{2 \zeta_F} \left \{ \pm (\frac{\mu_5}{\mu_0} +\zeta_F) + \frac{1}{2}(\frac{\mu_5}{\mu_0} +\zeta_F)^2 \right \} \nonumber \\
&\approx&  \pm \Delta \equiv \pm \frac{\pi}{2 \zeta_F} (\frac{\mu_5}{\mu_0} +\zeta_F) .
\end{eqnarray}
\end{widetext}

Expanding $1/\sinh(l \lambda(1 \pm \delta k/k_{F0}))$ and $e^{-l\lambda_D \delta k/k_{F0}}$ up to the first order in $\delta k/k_{F0} \ll 1$ as follows
\begin{eqnarray}
1/\sinh(x) &=& \frac{1}{x} - \frac{x}{6} + \frac{7x^3}{360} -... \\
e^{-x} &=& 1-x+\frac{x^2}{2} - ... ,
\end{eqnarray}
we keep $l=2$ components and sum SdH quantum oscillations from both chiral Fermi surfaces. As a result, we obtain
\begin{widetext}
\begin{eqnarray}
\sigma_{zz}^{osc} &=& \sigma_{zz}^{osc +} + \sigma_{zz}^{osc -}  \nonumber \\
&\approx&  C_1 B^{\frac{1}{2}}\frac{e^{-\lambda_D}}{\lambda} \left \{ (1-\frac{\delta k}{k_{F0}})(1-\lambda_D \frac{\delta k}{k_{F0}})\left( \cos(\Delta)\cos(\frac{\pi \hbar k_{F0}^2}{eB} + \frac{\pi}{4}) - \sin(\Delta)\sin(\frac{\pi \hbar k_{F0}^2}{eB}+ \frac{\pi}{4})\right) \right. \nonumber \\
&& \qquad \qquad \qquad +\left. (1+\frac{\delta k}{k_{F0}})(1+\lambda_D \frac{\delta k}{k_{F0}}) \left( \cos(\Delta)\cos(\frac{\pi \hbar k_{F0}^2}{eB} + \frac{\pi}{4}) + \sin(\Delta)\sin(\frac{\pi \hbar k_{F0}^2}{eB}+ \frac{\pi}{4})\right) \right \} \nonumber \\
&+& C_1 B^{\frac{1}{2}}\frac{e^{-2\lambda_D}}{2\sqrt{2} \lambda} \left \{(1-\frac{\delta k}{k_{F0}})(1-2\lambda_D \frac{\delta k}{k_{F0}}) \left (\cos(2 \Delta)\cos(\frac{2 \pi \hbar k_{F0}^2}{eB} + \frac{\pi}{4}) - \sin(2 \Delta) \sin (\frac{2 \pi \hbar k_{F0}^2}{eB}+ \frac{\pi}{4}) \right) \right. \nonumber \\
&& \quad \quad \qquad \qquad +\left. (1+\frac{\delta k}{k_{F0}})(1+2\lambda_D \frac{\delta k}{k_{F0}}) \left (\cos(2 \Delta)\cos(\frac{2 \pi \hbar k_{F0}^2}{eB} + \frac{\pi}{4}) + \sin(2 \Delta) \sin (\frac{2 \pi \hbar k_{F0}^2}{eB}+ \frac{\pi}{4}) \right) \right \} . \label{approxsigmatotal}
\end{eqnarray}
\end{widetext}
The procedure is essentially same for $\sigma_{xx}^{osc}$, not shown here.

\section{Origin of the double-peak structure in the SdH quantum oscillation and appearance of the kink structure in the Landau fan diagram}

The above longitudinal conductivity can be analyzed for three cases; two limiting cases and one intermediate case defined by a control parameter $\Delta = \frac{\pi}{2 \zeta_F} (\frac{\mu_5}{\mu_0} +\zeta_F)$. Two limiting cases will allow/forbid double peaks by Landau-level further splitting in the quantum oscillations whereas the intermediate parameter region discusses more general cases between such two limiting cases. We note that all oscillating parts of the conductivity will be normalized by $\sigma_{zz}^{osc} (\v{B}=0)$ later on.

\subsection{The limit of $\tan(\Delta) \rightarrow 0$}

\begin{figure}
\centering
\includegraphics[width=8cm]{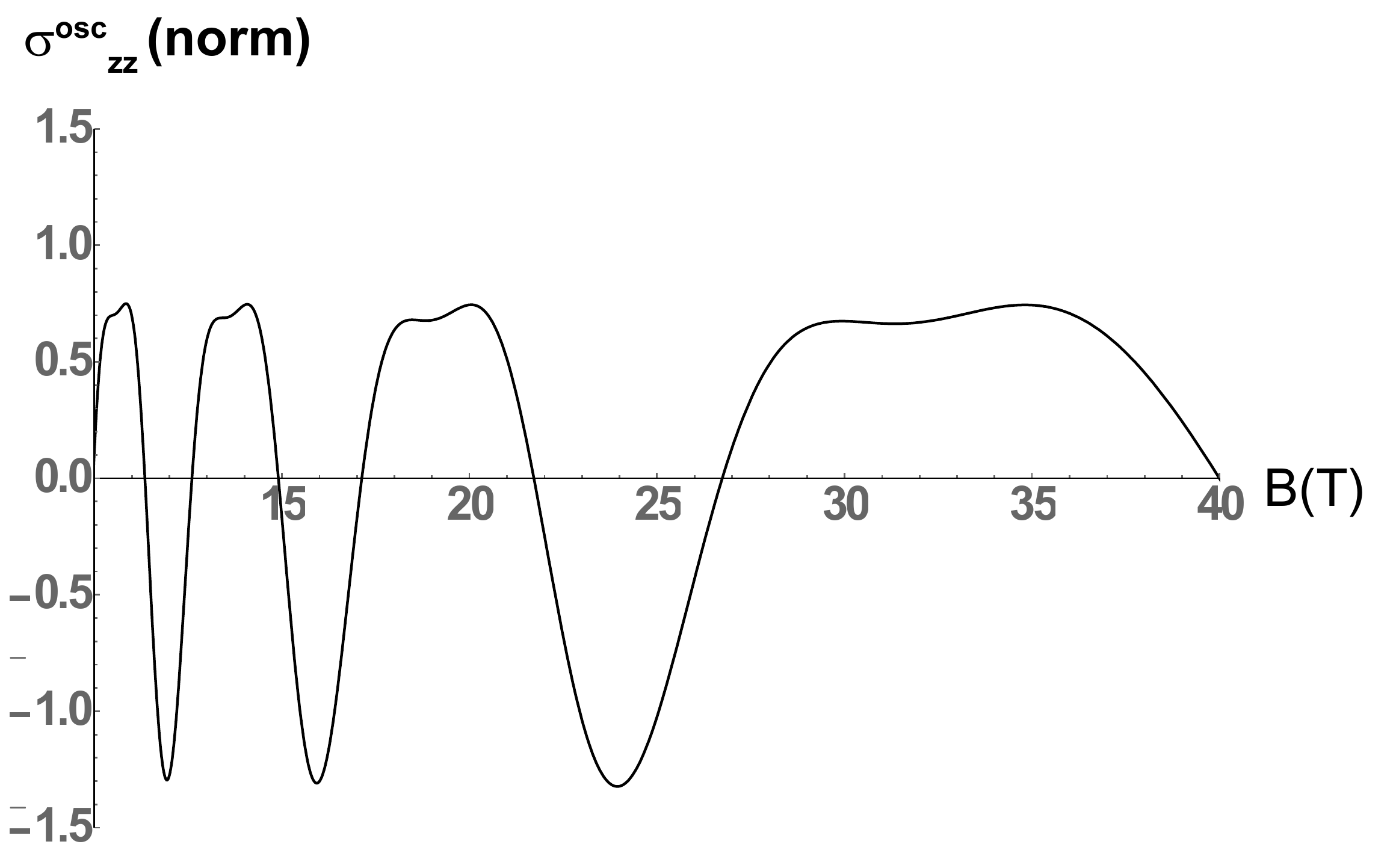}
\caption{The oscillating component of the longitudinal conductivity ($\sigma_{zz}^{osc}$) in the $\tan (\Delta) \rightarrow 0$ limit.} \label{TanD0}
\end{figure}

The first case we consider is when the effect for the sum of SdH oscillations from both chiral Fermi surfaces is minimized. This occurs when $\tan(\Delta) \rightarrow 0$, i.e., $\Delta / \pi = \frac{1}{2 \zeta_F} (\frac{\mu_5}{\mu_0} +\zeta_F) = m$, where $m$ is an integer. Then, the oscillating component of the conductivity is expressed as
\begin{widetext}
\begin{eqnarray}
\sigma_{zz}^{osc} &\approx&  \pm C_1 B^{\frac{1}{2}}\frac{e^{-\lambda_D}}{\lambda} \left \{ (1-\frac{\delta k}{k_{F0}})(1-\lambda_D \frac{\delta k}{k_{F0}}) \cos(\frac{\pi \hbar k_{F0}^2}{eB} + \frac{\pi}{4}) +(1+\frac{\delta k}{k_{F0}})(1+\lambda_D \frac{\delta k}{k_{F0}}) \cos(\frac{\pi \hbar k_{F0}^2}{eB} + \frac{\pi}{4}) \right \} \nonumber \\
&+& C_1 B^{\frac{1}{2}}\frac{e^{-2\lambda_D}}{2\sqrt{2} \lambda} \left \{ (1- \frac{\delta k}{k_{F0}})(1-2 \lambda_D \frac{\delta k}{k_{F0}}) \cos(\frac{2\pi \hbar k_{F0}^2}{eB} + \frac{\pi}{4}) +(1+\frac{\delta k}{k_{F0}})(1+2\lambda_D \frac{\delta k}{k_{F0}}) \cos(\frac{2 \pi \hbar k_{F0}^2}{eB} + \frac{\pi}{4}) \right \} \nonumber \\
&\approx& C_1 B^{\frac{1}{2}}\frac{2 e^{-\lambda_D}}{\lambda} \left ( \pm \cos (\frac{\pi \hbar k_{F0}^2}{e B} + \frac{\pi}{4}) + \frac{e^{-\lambda_D}}{2\sqrt{2}} \cos (\frac{2 \pi \hbar k_{F0}^2}{e B} + \frac{\pi}{4})  \right ).  \label{Delta0osc}
\end{eqnarray}
\end{widetext}
Recall that we keep all terms only in the first order of $\delta k/k_{F0}$. As shown in Eq. (\ref{Delta0osc}), there is no $\delta k$ term in this case. With the expansion of $1/\sinh(l \lambda(1 \pm \delta k/k_{F0}))$ and $e^{-l\lambda_D \delta k/k_{F0}}$ up to the first order in $\delta k/k_{F0}$, this equation is exactly the same as that of the conventional SdH oscillation in a metal. In this limit, the effect of the Fermi momentum change can be verified only when the order of the expansion is higher than the second order. Therefore, the SdH oscillation is almost the same as the conventional one and it is difficult to see double peaks by Landau level splitting. See Fig. \ref{TanD0}.

\subsection{The limit of $\tan(\Delta) \rightarrow \infty$}

On the other hand, the effect of the sum is maximized when $\tan(\Delta) \rightarrow \infty$, i.e., $\Delta / \pi = \frac{1}{2 \zeta_F} (\frac{\mu_5}{\mu_0} +\zeta_F) = \frac{1}{2} + m$, where $m$ is an integer. In this case, the oscillatory part of the longitudinal conductivity is given by
\begin{widetext}
\begin{eqnarray}
\sigma_{zz}^{osc} &\approx&  \pm C_1 B^{\frac{1}{2}}\frac{e^{-\lambda_D}}{\lambda} \left \{ -(1-\frac{\delta k}{k_{F0}})(1-\lambda_D \frac{\delta k}{k_{F0}}) \sin(\frac{\pi \hbar k_{F0}^2}{eB} + \frac{\pi}{4}) +(1+\frac{\delta k}{k_{F0}})(1+\lambda_D \frac{\delta k}{k_{F0}}) \sin(\frac{\pi \hbar k_{F0}^2}{eB} + \frac{\pi}{4}) \right \} \nonumber \\
&-& C_1 B^{\frac{1}{2}}\frac{e^{-2\lambda_D}}{2\sqrt{2} \lambda} \left \{ (1- \frac{\delta k}{k_{F0}})(1-2 \lambda_D \frac{\delta k}{k_{F0}}) \cos(\frac{2\pi \hbar k_{F0}^2}{eB} + \frac{\pi}{4}) +(1+\frac{\delta k}{k_{F0}})(1+2\lambda_D \frac{\delta k}{k_{F0}}) \cos(\frac{2 \pi \hbar k_{F0}^2}{eB} + \frac{\pi}{4}) \right \} \nonumber \\
&\approx& \pm C_1 B^{\frac{1}{2}}\frac{2e^{-\lambda_D}}{\lambda} (1+\lambda_D) (\frac{\mu_5}{\mu_0}+\zeta_F) \left ( \sin (\frac{\pi \hbar k_{F0}^2}{e B} + \frac{\pi}{4}) \mp \frac{e^{-\lambda_D}}{\mu_5/\mu_0 + \zeta_F}\frac{1}{2\sqrt{2}(1+\lambda_D)} \cos (\frac{2 \pi \hbar k_{F0}^2}{e B} + \frac{\pi}{4})  \right ) . \label{Deltainfosc}
\end{eqnarray}
\end{widetext}
Here, we keep all the terms in the first order of $\delta k/ k_{F0}$ again. The coefficients of sine and cosine functions are significantly modified to those of the previous case. Small factors from the numerator (Dingle factor $e^{-\lambda_D}$) and the denominator ($\mu_5/\mu_0 + \zeta_F$) are competing, so $l=2$ components for SdH oscillations (the second term in the last parenthesis in Eq. (\ref{Deltainfosc})) may survive in this limit. In particular, there is a special situation which always satisfies this condition ($\tan(\Delta) \rightarrow \infty$) in Weyl metals; An experimental situation of measuring transverse magnetoresistance. In this experimental set up, $\mu_5$ is always zero due to the orthogonality of $\v{E}$ and $\v{B}$, but there is the band-dispersion change due to the Berry curvature, and the $\zeta_F$ correction to the Fermi momentum exists. See Eq. (\ref{zetaFcorrection}). $\Delta$ is always $\pi/2$ in this case, which satisfies the second limit.

\begin{figure}
\centering
\includegraphics[width=8cm]{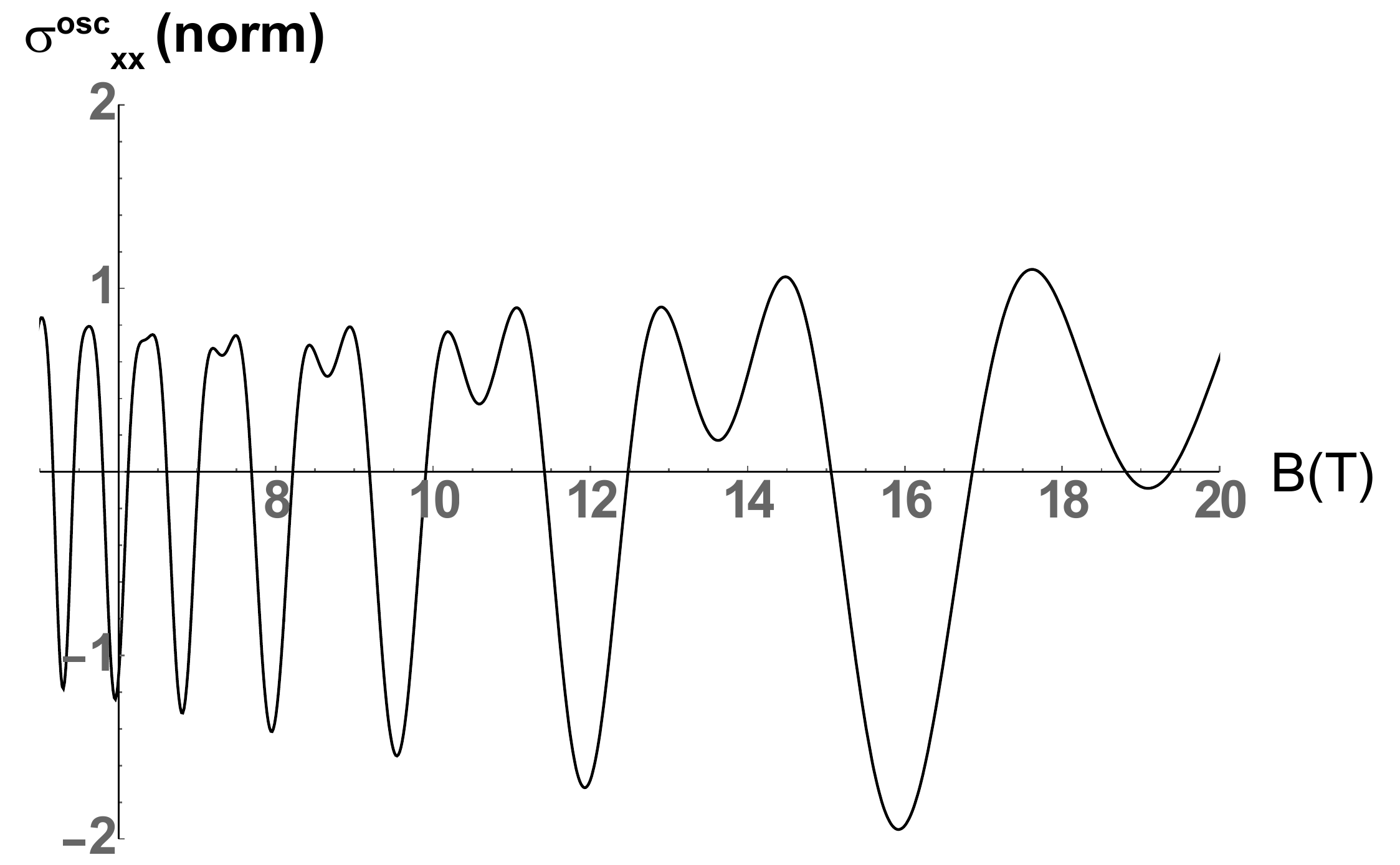}
\caption{The oscillating component of the longitudinal conductivity $\sigma_{xx}^{osc}$ in the $\tan (\Delta) \rightarrow \infty$ limit} \label{TanDinf}
\end{figure}

To verify this statement, we consider the transverse oscillatory components, given by
\begin{eqnarray}
\sigma_{xx}^{osc} &\approx& \pm C_1 B^{\frac{1}{2}}\frac{e^{-\lambda_D}}{\lambda}\frac{(1+\lambda_D)}{9 + (2 k_F v_F \zeta_F \tau)^2 } \times \nonumber \\
&& \left ( \sin (\frac{\pi \hbar k_{F0}^2}{e B} + \frac{\pi}{4}) \mp \frac{e^{-\lambda_D}}{\zeta_F}\frac{\cos (\frac{2 \pi \hbar k_{F0}^2}{e B} + \frac{\pi}{4})}{\sqrt{2}(1+\lambda_D)} \right ) . \nonumber \\ \label{xxosc}
\end{eqnarray}
This result is quite similar to that of Eq. (\ref{Deltainfosc}). Because of the competition between the numerator (Dingle factor $e^{-\lambda_D}$) and the denominator ($\zeta_F$) in the second term of Eq. (\ref{xxosc}), one can expect double peaks in SdH quantum oscillations. See Fig. \ref{TanDinf}.

\subsection{General experimental setup}

In the general case, $\tan (\Delta)$ would be in a range of $0 < \tan (\Delta) < \infty$. We can consider this intermediate regime as $\cos (\Delta) > (\frac{\delta k}{k_{F0}})$ $\&$ $\sin (\Delta) \neq 0$. Keeping all terms in the first order of $\delta k/k_{F0}$ in Eq. (\ref{approxsigmatotal}), we get an approximate oscillatory expression of the conductivity as
\begin{widetext}
\begin{eqnarray}
\sigma^{osc}_{zz} &\approx& \frac{2 C_1 B^{\frac{1}{2}} e^{-\lambda_D}}{\lambda} \left \{ \sqrt{\cos^2(\Delta) + (1+\lambda_D)^2 (\frac{\delta k}{k_{F0}})^2 \sin^2(\Delta)} \cos (\frac{\pi \hbar k_{F0}^2}{eB} + \frac{\pi}{4} - \phi_1)  \nonumber \right. \\
&& \left. + \frac{e^{-\lambda_D}}{2\sqrt{2}} \sqrt{\cos^2(2\Delta) + (1+2\lambda_D)^2 (\frac{\delta k}{k_{F0}})^2 \sin^2(2\Delta)} \cos (\frac{2 \pi \hbar k_{F0}^2}{eB} + \frac{\pi}{4} - 2\phi_2) \right \} \nonumber \\
&\approx& \frac{2 C_1 B^{\frac{1}{2}} e^{-\lambda_D}}{\lambda} \cos (\Delta) \left \{ \cos (\frac{\pi \hbar k_{F0}^2}{eB} + \frac{\pi}{4} - \phi_1) + \frac{e^{-\lambda_D}}{2\sqrt{2}} \frac{\cos (2\Delta)}{\cos (\Delta)} \cos (\frac{2 \pi \hbar k_{F0}^2}{eB} + \frac{\pi}{4} - 2\phi_2) \right \} , \label{Deltaintermediateosc}
\end{eqnarray}
\end{widetext}
where $\tan \phi_1 = (1+\lambda_D) \frac{\delta k}{k_{F0}}\tan (\Delta)$ and $\tan 2\phi_2 = (1+2\lambda_D) \frac{\delta k}{k_{F0}}\tan (2\Delta)$. One can easily check out that $\Delta \rightarrow 0$ and $\Delta \rightarrow \pi/2$ correspond to the first and second limits, respectively.

\begin{figure}
\centering
\includegraphics[width=8cm]{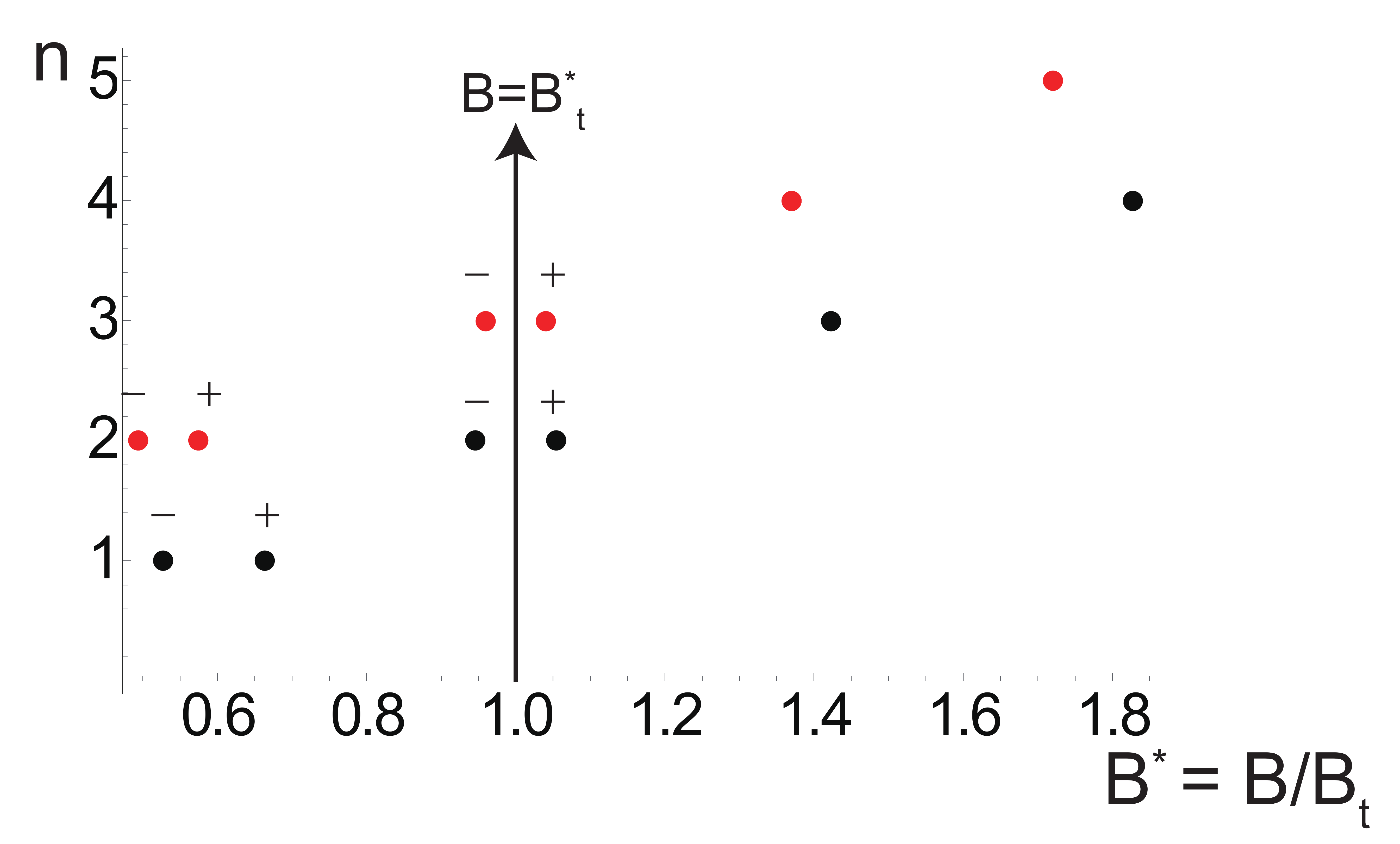}
\caption{Landau fan diagram with Landau level further splitting with a pair of chiral Fermi surfaces. $+$ $(-)$ indicates each chirality. The upper red (Ca$_3$As$_2$ \cite{Cd3As2Cao}) and the lower black (ZrTe$_5$ \cite{ZrTe5Zheng}) dots came from different samples. Here, we normalized the magnetic field (B) by each threshold field $B_t$. See the text for more details.} \label{fitting}
\end{figure}

\begin{figure*}
\centering
\includegraphics[width=16cm]{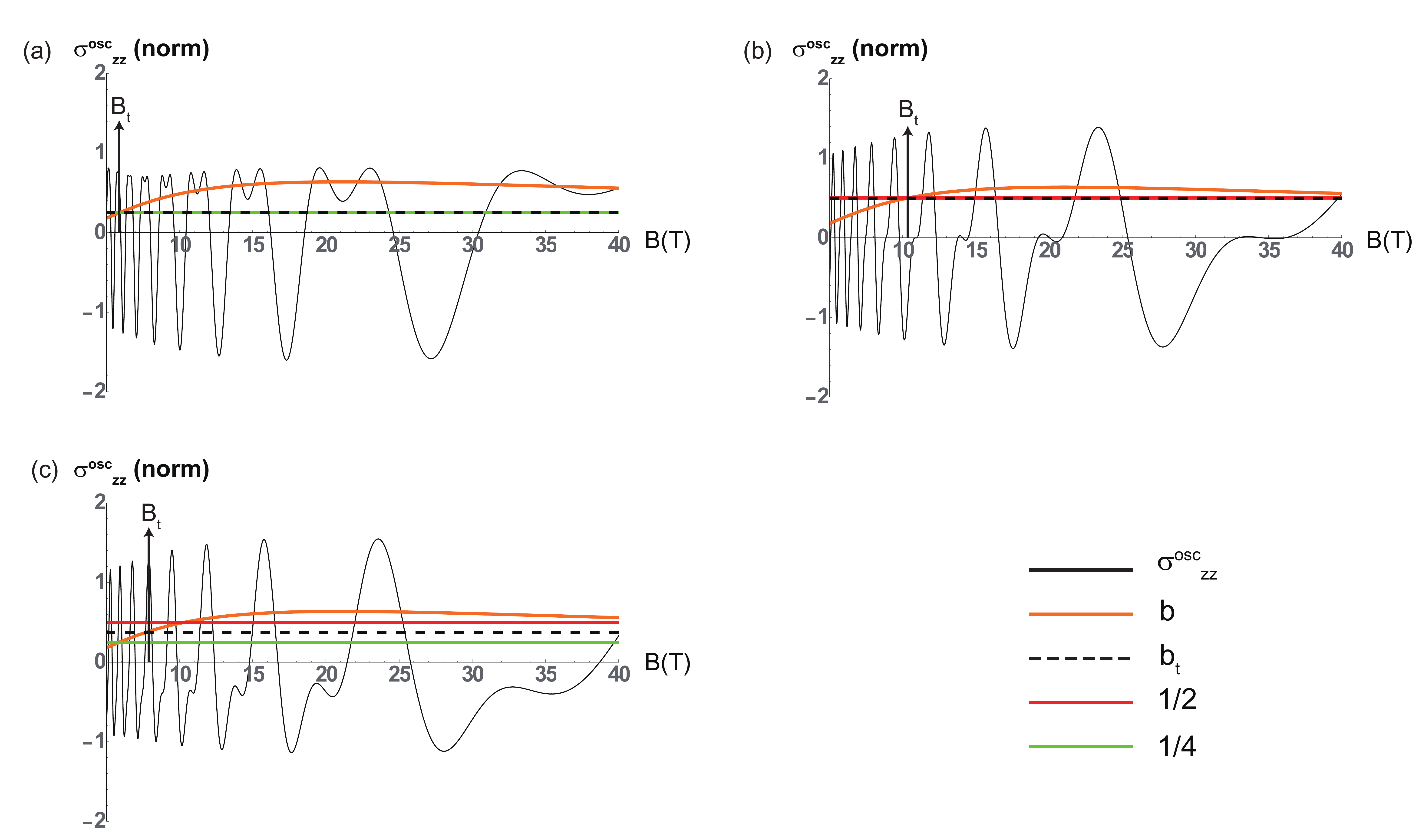}
\caption{Oscillating components of the longitudinal conductivity $\sigma_{zz}^{osc}$ in general cases. The threshold magnetic field $B_t$ is determined by the dimensionless parameter $b$ (orange line) with the condition $b=b_t$. Double peaks in quantum oscillations start to appear when the external magnetic field exceeds the threshold value, i.e., $B>B_t$. The threshold value $b_t$ (dashed black line) always exists between $\frac{1}{4}$ (green line) and $\frac{1}{2}$ (red line). Three different cases with various values of $\alpha$ and $\beta$ are shown. a. $\alpha - \beta = \frac{m \pi}{2}$ where $b_t = \frac{1}{4}$ is given as a minimum. b. $\alpha - \beta = \frac{m \pi +\pi/2}{2}$ where $b_t = \frac{1}{2}$ is given as a maximum. c. $\alpha - \beta = \pi/8$ where $b_t=3/8$ is between the minimum and maximum values. See the text for more details.} \label{intermediate}
\end{figure*}

\subsection{Analyzing each limit with the threshold field $B_t$}

Let us consider a function $f(x)$ with arbitrary phases $\alpha$ and $\beta$, given by
\begin{eqnarray}
f(x) &=& \cos(1/x -\alpha)+ b \cos(2/x - 2\beta) \label{fx}.
\end{eqnarray}
This function has the same form as Eqs. (\ref{Delta0osc}) $\sim$ (\ref{Deltaintermediateosc}), where $b$ corresponds to $\frac{e^{-\lambda_D}}{2\sqrt{2}}$ in the $\tan(\Delta) \rightarrow 0$ limit, $\frac{e^{-\lambda_D}}{\mu_5/\mu_0 +\zeta_F} \frac{1}{2\sqrt{2}(1+\lambda_D)}$ in the $\tan(\Delta) \rightarrow \infty$ limit, and $\frac{e^{-\lambda_D}}{2\sqrt{2}} \frac{\cos (2\Delta)}{\cos (\Delta)}$ in the intermediate regime of $0 < \tan(\Delta) < \infty$, respectively. We also point out that $1/x = \frac{\pi \hbar k_{F0}^2}{eB}$, $\alpha = \phi_1-\pi/4$, and $\beta=\phi_2-\pi/8$. When $|b|$ is small (note that $b$ is a function of the external magnetic field), the first term in Eq. (\ref{fx}) is dominating. We only see the oscillation peaks with $1/x$ period. However, when the external magnetic field $B$ is larger than a threshold field $B_t$, $|b|$ becomes larger than a threshold value $b_t$ and the $2/x$ period term starts to show its effect. Here, the threshold value $b_t$ can be defined by the existence of two multiple root in the vanishing first derivative of $f(x)$, i.e., given by $f'(x) = f''(x) = 0$. On the other hand, the vanishing double derivative of $f(x)$ sometimes appears in the absence of the vanishing first derivative of $f(x)$. To avoid this possibility in determining $b_t$, we suggest to consider only the first derivative of $f(x)$, where its vanishing condition gives a solution, the period of which differs from the existing one. In this respect the threshold value may be regarded to be qualitative. For the $B>B_t$ region in any Weyl metals, double peaks in SdH quantum oscillations have to occur due to the Landau level further splitting as shown in Fig. \ref{fitting}. Even though every sample has its different threshold limit, analyzing the function form of $f(x)$, one can easily find the threshold field in a physical sense.

The threshold value $b_t$ is $1/4$ (minimum) when $\alpha - \beta = \frac{m \pi}{2}$ whereas $b_t$ is $1/2$ (maximum) for $\alpha - \beta = \frac{(m+1/2) \pi}{2}$. These special values are determined by $f'(x) = f''(x) = 0$. It is not simple to express $b_t$ as an analytic form for arbitrary $\alpha$ and $\beta$ as discussed above, but $b_t$ always exists in the range of $1/4 < b_t \leq 1/2$. We show three examples of $\alpha - \beta = \frac{m\pi}{2}, \frac{m\pi+\pi/2}{2}, \text{ and } \frac{\pi}{8}$ in Fig. \ref{intermediate}. We find the threshold values of $b_t$ and $B_t$ in a numerical way when $\alpha$ and $\beta$ are arbitrarily given (whenever $b_t$ is given as a number, $B_t$ can be found by solving $b=b_t$ in a numerical way). When $|b|$ is much larger than $1/2$, one can expect to see sufficiently big oscillation peaks with the $2/x$ period. In the first limit ($\tan (\Delta) \rightarrow 0$), it is extremely hard to see the $2/x$ period oscillation peaks because the maximum value of $b$ is not sufficiently large ($\frac{1}{2\sqrt{2}}$). Even if the $2/x$ period oscillation peaks exist, the amplitudes of them are extremely small compared to that of the $1/x$ period oscillations. This is the reason why double peaks are rare in conventional metals. On the other hand, in Weyl metals, the amplitude of $|b|$ can be arbitrarily tuned depending on $\mu_5$ and $\zeta_F$ and the system can go to the second limit ($\tan (\Delta) \rightarrow \infty$). In other words, tuning $\mu_5$ with the applied electric field, one can manipulate the double-peak condition in Weyl metals. The easiest way to control the $\Delta$ parameter in experiments might be changing the angle between the external electric field $\v{E}$ and the external magnetic field $\v{B}$. One can manipulate the $\Delta$ parameter from $\pi/2$ (at $\v{E} \perp \v{B}$ ) to a certain maximum value (at $\v{E} \parallel \v{B}$) by changing the angle. Such tuning of double peaks (i.e., tuning the Landau level splitting effect in quantum oscillations) is possible only in Weyl metals thanks to the chiral charge pumping. The tuning conditions of double peaks are summarized in Table \ref{parameters} with equations of $|b|$ for certain conditions.

\begin{table}
\begin{tabular}{c @{$\quad$ $\quad$ } c @{$\quad$ $\quad$ }}
\\
\hline \hline $\Delta$ & Condition for the $2/x$ period peaks \\
\hline \\
Arbitrary $\Delta$ & $\abs{b} = \frac{e^{-\lambda_D}}{2\sqrt{2}} \frac{\cos (2\Delta)}{\cos (\Delta)} > b_t $ \\
$\tan{\Delta} \rightarrow 0$ & $\abs{b}= \frac{e^{-\Lambda_D}}{2\sqrt{2}}>b_t$ \\
$\tan{\Delta} \rightarrow \infty$ & $\abs{b} = \abs{\frac{e^{-\lambda_D}}{2\sqrt{2}(\mu_5/\mu_0+\zeta_F)(1+\lambda_D)}\zeta_F} > b_t$ \\
\\
\hline \hline \\
\end{tabular}
\caption{Condition for the $2/x$ period peaks in SdH oscillations depending on the parameter $\Delta$.} \label{parameters}
\end{table}

\subsection{$\mu_5$ measurement}

Based on the above analysis, we suggest a method to obtain the $\mu_5$ value experimentally. Observing quantum oscillations in both transverse and longitudinal directions, one may evaluate $\mu_5$ of the system approximately with the following experiment. First, measure the double peaks in the SdH oscillations for the transverse direction. One might get the oscillating amplitudes of the $1/x$ and $2/x$ components using the Fourier transform. From Eq. (\ref{xxosc}), we know that the ratio of oscillating amplitudes between the $1/x$ and $2/x$ components should be given as $b_{\perp}=\frac{e^{-\lambda_D}}{\sqrt{2}(1+\lambda_D)\zeta_F}$. Comparing it to the measured one, the value of $\lambda_D$ can be evaluated ($\zeta_F$ is given by the oscillating period as usual). Same process in the longitudinal direction can give the information of $\Delta = \frac{\pi}{2}(\frac{\mu_5}{\mu_0}/\zeta_F + 1)$. Controlling the amplitude of $\v{E}$ during the experiment, one can find the second-limit condition ($\tan (\Delta) \rightarrow \infty$) by observing maximized amplitudes of double peaks. In this limit, one can use Eq. (\ref{Deltainfosc}) with the ratio of $b_{\parallel}=\frac{e^{-\lambda_D}}{\mu_5/\mu_0 + \zeta_F} \frac{1}{2\sqrt{2}(1+\lambda_D)}$. Even if finding the second limit is not successful, one can use Eq. (\ref{Deltaintermediateosc}) with the ratio of $b_{\parallel}=\frac{e^{-\lambda_D}}{2\sqrt{2}} \frac{\cos (2\Delta)}{\cos (\Delta)}$ in the intermediate region. From two equations with experimentally given $b_{\parallel}$ and $b_\perp$, one can find the value of $\Delta$ and $\lambda_D$. $\Delta$ immediately gives the value $\mu_5/\mu_0$.

We introduce one more method of measuring the $\mu_5$ directly. Measure the SdH oscillations at fixed $\v{E}$. Repeat this measurement for various amplitudes of $\v{E}$. Then, double peaks will appear when $\Delta = m\pi + \pi/2$ and disappear when $\Delta = m\pi$. It means that the double peaks appear when $\mu_5/\mu_0 = (2n) \zeta_F$, whereas they disappear when $\mu_5/\mu_0 = (2n+1) \zeta_F$. Let us define the repeating period of $E$ as $E_p$. Then, from Table \ref{parameters}, we obtain
\begin{eqnarray}
\frac{\mu_5(E_p, B)}{\mu_0} &=& 2 \zeta_F(B) \nonumber \\
\frac{\hbar v_F a \v{E} \cdot \v{B}}{\hbar v_F k_{F0}} &=& 2 \frac{eB}{2\hbar k_{F0}^2} \nonumber \\
\therefore a &=& \frac{e}{\hbar k_{F0} E_p} . \label{getmu5}
\end{eqnarray}
Therefore, one can find the coefficient in front of $\v{E} \cdot \v{B}$, which indicates the value of $\mu_5$ from Eq. (\ref{getmu5}).

\section{Summary}

In this paper, we investigated how the Landau-level further splitting can arise in a time-reversal symmetry-broken Weyl metal phase. In particular, we verified when a double-peak structure appears in the SdH quantum oscillations, responsible for a kink structure in the Landau fan diagram. It turns out that (i) Berry-curvature induced orbital magnetic moments give rise to chirality-dependent dispersion relations and (ii) chiral charge pumping effects cause an effective chiral chemical potential through the dissipationless current channel of the bulk sample. As a result, the area of each chiral Fermi surface becomes different as long as applied electric and magnetic fields satisfy a physical condition that we discussed in the main text.

We would like to emphasize that controlling the double-peak structure by tuning external $\v{E}$ and $\v{B}$ fields is only possible in Weyl metals because the Landau-level further splitting is governed by two different factors mentioned above. The other crucial point is that direct evaluations for the chiral chemical potential $\mu_5$ is possible by tuning external $\v{E}$ and $\v{B}$ fields.


%
%

\begin{acknowledgments}
K.-S. Kim was supported by the Ministry of Education, Science, and Technology (NRF-2021R1A2C1006453 and NRF-2021R1A4A3029839) of the National Research Foundation of Korea (NRF). We appreciate helpful discussions with H.-J. Kim and M. Sasaki.
\end{acknowledgments}

\section*{Appendix}
\setcounter{figure}{0}
\setcounter{equation}{0}
\renewcommand{\theequation}{A \arabic{equation}}

In this appendix, we solve the Boltzmann equation to obtain $\sigma_{xx}$ (transverse conductivity in the $x$ direction, which is perpendicular to the external magnetic field along the $z$ direction) of the Weyl metal system in details. We note that Ref. \cite{sigmaz} has already shown how to obtain $\sigma_{zz}$ with basically an identical method.

\subsection{Conductivity $\sigma_{ab}$ from current density $\v{j}$}

Current density in a Weyl metal phase is expressed as \cite{RMP82}
\begin{equation}
\v{j} = -2e \int_{\text{BZ}} f(\v{x},\v{k},t) \left [ \v{v_k} + \frac{e}{\hbar} (\v{v_k} \cdot \boldsymbol{\Omega})\v{B} + \frac{e}{\hbar} \v{E} \times \boldsymbol{\Omega} \right ] , \label{currentdensity}
\end{equation}
where $f(\v{x},\v{k},t)$ is the distribution function, $\v{v_k}$ is the group velocity, and $\boldsymbol{\Omega}$ is the Berry curvature in the momentum space. BZ indicates that the integration range is limited in the first Brillouin zone.

In the linear response regime, the distribution function is given by
%
%
\begin{equation}
f(\v{x},\v{k},t) = f_0(\varepsilon) + e \pd{f_0}{\varepsilon} \v{E} \cdot \v{g} + O(\v{E}^2) ,
\end{equation}
where $\v{g}$ is a near-equilibrium distribution function, determined by the Boltzmann equation.

Inserting this expression into Eq. (\ref{currentdensity}), we obtain the conductivity tensor as
\begin{eqnarray}
\sigma_{ab} &=& -2 e^2 \int \pd{f_0}{\varepsilon} g_b \left (\v{v_k} + \frac{e}{\hbar} (\v{v_k} \cdot \boldsymbol{\Omega}) \v{B} \right )_a \frac{d^3 k}{(2 \pi)^3} \nonumber \\
&& + \frac{2 e^2}{\hbar} \varepsilon_{abc} \int \Omega_c(\v{k}) f_0(\varepsilon) \frac{d^3 k}{(2 \pi)^3}, \\
&=& -2e^3 \sum_{\chi = \pm} \pd{f_0}{\varepsilon} g_b v_k (\hat{\v{k}} + \chi \zeta_k \hat{z})_a \frac{d^3 k}{(2 \pi)^3}. \label{sigmaab}
\end{eqnarray}
Here, we assumed an isotropic case for the last equality, where the Berry curvature $\boldsymbol{\Omega}(\v{k})$ is
\begin{equation}
\boldsymbol{\Omega}(\v{k}) = \chi \frac{\hat{\v{k}}}{2 k^2}.
\end{equation}

\subsection{Boltzmann equation}
\setcounter{figure}{0}
\setcounter{equation}{0}
\renewcommand{\theequation}{B \arabic{equation}}

To obtain $\sigma_{xx}$, we find $g_x$, governed by the following Boltzmann equation \cite{RMP82}
\begin{eqnarray}
&& \left [ \Upsilon (\partial_t + i \omega) - \frac{e}{\hbar} (\v{v_k} \times \v{B}) \cdot \boldsymbol{\nabla_{\v{k}}} \right ] \v{g} \nonumber \\
&=& \v{v_k} + \frac{e}{\hbar} (\v{v_k} \cdot \boldsymbol{\Omega}) \v{B} + \int_{BZ}\frac{d^3 k'}{(2 \pi)^3} (\Upsilon' \omega_{\v{k}' \rightarrow \v{k}} \Upsilon) (\v{g}' - \v{g}), \nonumber \\ \label{Boltzmann}
\end{eqnarray}
where $\Upsilon = 1+ \frac{e}{\hbar} \v{B} \cdot \boldsymbol{\Omega}(\v{k})$ is the phase-space volume factor. Here, we assume elastic scattering with a weak and short-range impurity potential. Then, the transition rate $\omega_{\v{k}' \rightarrow \v{k}}$ is given by
\begin{eqnarray}
\omega_{\v{k}' \rightarrow \v{k}} = \frac{3}{2 \nu(\varepsilon) \tau(\varepsilon)} (1+ \hat{\v{k'}} \cdot \hat{\v{k}}) \delta (\varepsilon-\varepsilon'), \label{omega}
\end{eqnarray}
where $\nu(\varepsilon)$ is the density of states at the energy $\varepsilon$ without external magnetic fields.

Incorporating Eq. (\ref{omega}) into Eq. (\ref{Boltzmann}), we obtain a self-consistent equation for $g_x$ as follows
\begin{eqnarray}
&& (i\omega \Upsilon - \frac{e}{\hbar} (\v{v_k} \times \v{B}) \cdot \boldsymbol{\nabla_{\v{k}}})g_x - v_x(k) \nonumber \\
 &=& \frac{3 \Upsilon}{16 \pi^3} \int d^3k' \Upsilon' (g_x'-g_x)\frac{1+\v{\hat{k}}' \cdot \v{\hat{k}}}{\nu(\varepsilon)\tau(\varepsilon)}\delta(\varepsilon - \varepsilon'),
\end{eqnarray}
where $\zeta_k = \frac{e B}{2 \hbar k^2}$ is the dimensionless length scale as mentioned in the text. Note that we are considering a stationary solution, so we are dealing with a time independent solution.

With an azimuthal symmetry, we assume the following ansatz of $g_x$ in the spherical coordinate as
\begin{eqnarray}
g_x(\theta, \phi) = \sum_m b_m(\theta) e^{i m \phi}. \label{gx}
\end{eqnarray}
Due to the $e^{i m \phi}$ term in the ansatz, all terms of $m>2$ disappear by the azimuthal-angle ($\phi$) integration. The resulting self-consistent equation of $b_m$ with the spherical coordinate reads
\begin{eqnarray}
&& (i\omega \Upsilon + 2 k \zeta_k v_k \pd{}{\phi})g_x(\theta, \phi) - v_k \sin{\theta} \cos{\phi} \nonumber \\
&=& \frac{3 \Upsilon}{16 \tau} \int d\theta' \Upsilon' \sin\theta' [ 2(b'_0-g_x(\theta, \phi))(1+\cos\theta\cos\theta') \nonumber \\
&& +(b'_1+b'_{-1})\sin\theta'\sin\theta\cos\phi+i(b'_1-b'_{-1})\sin\theta'\sin\theta\sin\phi ], \nonumber \\
\label{}
\end{eqnarray}
where $b'_i = b_i(\theta')$ and $\Upsilon' = \Upsilon(\theta')$.

Comparing all terms between the left and right sides of Eq. (\ref{}) after the polar-angle ($\theta'$) integration, we obtain $b_m(\theta)$ terms as follows
\begin{eqnarray}
b_0(\theta) &=& \frac{\frac{3 \Upsilon}{8 \tau} \int d\theta' \Upsilon' \sin\theta'b'_0(1+\cos\theta\cos\theta')}{[i \omega \Upsilon + \frac{3\Upsilon}{4 \tau}(1+\frac{1}{3}\chi \zeta_k \cos\theta)]} \nonumber \\
 &=& \frac{\alpha_0 +\beta_0\cos\theta}{[\frac{8\omega\tau}{3}i + 2(1+\frac{1}{3}\chi \zeta_k \cos\theta)]}, \\
b_{\pm 1}(\theta) &=& \frac{\sin\theta (v_k+ \Upsilon u_{\pm})}{\Upsilon[i \omega + \frac{3}{4 \tau}(1+\frac{1}{3}\chi \zeta_k \cos\theta)] \pm 2k\zeta_k v_k i}, \nonumber \\
\end{eqnarray}
where the constants of $\alpha_0$, $\beta_0$, and $u_{\pm}$ are
\begin{eqnarray}
\alpha_0 &\equiv& \int d\theta' \Upsilon' \sin\theta'b_0(\theta'), \\
\beta_0 &\equiv& \int d\theta' \Upsilon' \sin\theta' \cos\theta' b_0(\theta'), \\
u_{\pm} &\equiv& \frac{3}{16 \tau}\int d\theta' \Upsilon' \sin^2\theta'b_{\pm 1}(\theta').
\end{eqnarray}

In the semi-classical limit where a large number of Landau levels are filled with weak external magnetic fields, $\frac{1}{2} k_F^2 l_B^2 = 1/2\zeta_F \gg 1$ should be satisfied, where $l_B \equiv \sqrt{\frac{\hbar}{eB}}$ is the magnetic length. Therefore, $\zeta_k = 1/(k^{2} l_B^2) \ll 1$ is satisfied in the vicinity of the Fermi surface. In this limit, the three constants $\alpha_0$, $\beta_0$, and $u_{\pm}$ can be approximated as
\begin{eqnarray}
\alpha_0 &\approx& \sqrt{\frac{1}{2}}\left(1-\frac{1}{15}\chi \zeta_k + \frac{1}{300}(\zeta_k)^2\right), \\
\beta_0 &\approx& \frac{9}{2}(1+ \frac{2}{\chi \zeta_k} + \frac{2\chi \zeta_k}{27})\alpha_0, \\
 u_{\pm} &\approx& -\frac{i v_k}{2 (-i \pm 4 \tau k\zeta_k v_k + 2 \tau \omega)} \nonumber \\
 && \pm \frac{8 (\tau k\zeta_k v^2 + i \omega \tau^2k\zeta_k v^2) \zeta_k^2}{
 5 (-i \pm 4 \tau k\zeta_k v + 2 \tau \omega)^2 (-3 i \pm 8 \tau k\zeta_k v + 4 \tau \omega)}. \nonumber \\
\end{eqnarray}

\subsection{Evaluation of $\sigma_{xx}$}

\setcounter{figure}{0}
\setcounter{equation}{0}
\renewcommand{\theequation}{C \arabic{equation}}

Inserting the near-equilibrium distribution function with the presence of weak external magnetic fields into Eq. (\ref{sigmaab}), we obtain $\sigma_{xx}$ as
\begin{widetext}
\begin{eqnarray}
\sigma_{xx} &=& -\frac{e^2}{8 \pi^3}\sum_n \iiint \delta(n-\frac{1}{4}(\sin^2\theta/\zeta_k - 2 \chi \cos \theta))\pd{f_0}{\epsilon}v_k \sin \theta \cos \phi k^2\sin \theta g_x  d\phi d\theta dk \nonumber \\
&=& \frac{-e^2}{8 \pi^2} \sum_l \iint e^{2 \pi l n i}\pd{f_0}{\epsilon} v_k k^2 \sin^2 \theta [b_1(\theta) + b_{-1}(\theta)] d\theta dk \nonumber \\
&=& \frac{-e^2}{8 \pi^2}\sum_l \int_0^\infty \pd{f_0}{\epsilon} v_k k^2 e^{i\frac{\pi l}{2}(\zeta_k^{-1}+\zeta_k)}\int_0^{\frac{\pi}{2}}e^{\frac{-\pi l i}{2\zeta_k}(\cos \theta + \chi \zeta_k)^2} \sin^2 \theta [b_1(\theta) + b_{-1}(\theta)]d\theta dk. \label{sigmaxxbm}
\end{eqnarray}
\end{widetext}
Note that $\delta(n-\frac{1}{4}(\sin^2\theta/\zeta_k - 2 \chi \cos \theta))$ term comes from the discreteness of the Fermi surface due to the Bohr-Sommerfeld quantization condition.

To go further with this expression, we resort to the poisson re-summation formula for the second line
\begin{eqnarray}
\sum_{n=-\infty}^{\infty} \delta(x-n) = \sum_{l=-\infty}^{\infty}e^{i2\pi l x},
\end{eqnarray}
where $x=\frac{1}{4}(\sin^2\theta/\zeta_k - 2 \chi \cos \theta$). In Eq. (\ref{sigmaxxbm}), integrating over the momentum $k$ can be easily treated because of the $\pd{f_0}{\epsilon} \approx -\frac{\delta(k-k_F)}{\hbar v_F}$ term. On the other hand, an exact integration over the polar angle ($\theta$) is not trivial because of the complicated form of $b_{\pm 1} (\theta)$.

Expanding the above expression up to the second order of $\zeta_k$ in the small $\zeta_k$ limit, the polar-angle integral can be performed as
\begin{eqnarray}
&&\int_0^{\frac{\pi}{2}}e^{\frac{-\pi l i}{2\zeta_k}(\cos \theta + \chi \zeta_k)^2} \sin^2 \theta [b_1(\theta) + b_{-1}(\theta)]d\theta \nonumber \\
&\approx& (C_0 + C_0'\zeta_k^2)Q_0 + C_1\zeta_k Q_1 + (C_2 + C_2'\zeta_k^2)Q_2 \nonumber \\
&& + C_3 \zeta_k Q_3 + C_4 \zeta_k^2 Q_4  \label{polarangleintegral}
\end{eqnarray}
where $C_i$ and $C_i'$ are functions of $k$ but independent of $\theta$, and $Q_m$ are defined as
\begin{eqnarray}
Q_m &\equiv& \int_{-1 + \chi \zeta_k}^{1+\chi \zeta_k}y^m e^{-\frac{-i\pi l}{2\zeta_k}y^2}dy, \nonumber \\
y &\equiv& \cos \theta + \chi \zeta_k. \nonumber
\end{eqnarray}

Considering an energy $\varepsilon$ window near the chemical potential $\mu$, following approximations should be valid on Eq. (\ref{polarangleintegral})
\begin{eqnarray}
\zeta_k &=& \sum_{n=0}^\infty \frac{\partial_{\varepsilon}^n\zeta_k(\varepsilon = \mu)}{n!}(\varepsilon - \mu)^n \nonumber \\
&\approx& \zeta_F(1-\frac{2(\varepsilon-\mu)}{\varepsilon_F}), \\
\zeta_k^{-1} + \zeta_k &\approx& \zeta_F^{-1}(1 + \zeta_F^2) + \frac{2(\varepsilon-\mu)}{\varepsilon_F \zeta_F}(1-\zeta_F^2).
\end{eqnarray}

Integrating over $\theta$ with this approximation, we find that the $Q_1$ term vanishes and $Q_m$ terms for $m>2$ are in higher orders than $O(\zeta_k^2)$. Resulting integrals for $Q_0$ and $Q_2$ are given by
\begin{eqnarray}
Q_0 &=& \sqrt{\frac{2 \zeta_k}{i l}} + \frac{2 \zeta_k (-1)^l}{\pi l} \frac{e^{-i \frac{\pi l}{2}(\zeta_k^{-1}+\zeta_k)}}{1-\zeta_k^2}, \nonumber \\
Q_2 &=& -\frac{e^{\frac{i\pi}{4}}}{\pi} \sqrt{\frac{2\zeta_k^3}{l^3}} + \frac{2 \zeta_k (-1)^l}{\pi l} e^{-i \frac{\pi l}{2}(\zeta_k^{-1}+\zeta_k)}(i+\frac{2 \zeta_k}{\pi l}). \nonumber
\end{eqnarray}
Therefore, $\sigma_{xx}$ up to the order of $\zeta_k^2$ is
\begin{eqnarray}
\sigma_{xx} &\approx&  \sigma_{xx}^{l=0} -\frac{e^2}{8 \pi^2} \sum_{l=1}^{\infty} (v_F k_F^2) \int_0^\infty \pd{f_0}{\epsilon}e^{\frac{\pi l i}{2}(\zeta_k^{-1}+\zeta_k)} \nonumber \\
&& \left \{\frac{2\zeta_F(-1)^l}{\pi l}[\frac{1}{1-\zeta_k^2}C_0+(i+2\frac{\zeta_F}{\pi l})C_2] \right . \nonumber \\
&& \left . +\sqrt{\frac{2\zeta_k}{li}}[C_0-C_2\frac{\zeta_ke^{\frac{\pi i}{2}}}{l\pi}] \right \} dk, \nonumber \\ \label{sigmaxxc0c2}
\end{eqnarray}
where $C_0$ and $C_2$ are
\begin{eqnarray}
C_0 &\approx& \sum_{j = \pm} \frac{4 \tau}{a_j}(v_F + u_j), \label{C0}\\
C_2 &\approx& -\sum_{j = \pm} \frac{4 \tau}{a_j}(v_F + u_j), \label{C2}\\
\text{with} \nonumber \\
a_\pm &\equiv& 3 + 4 \omega \tau i \pm 8 k_F v_F \tau \zeta_F, \nonumber \\
 u_{\pm} &\approx& -\frac{i v_k}{2 (-i \pm 4 k v_k \tau \zeta_k + 2 \tau \omega)} \nonumber \\
&&\pm \frac{8 (k v_k \tau \zeta_k v_k + i k v_k \tau \zeta_k v_k\tau\omega) \zeta_k^2}{
 5 (-i \pm 4 kv_k\tau \zeta_k  + 2 \tau \omega)^2 (-3 i \pm 8 k v_k \tau \zeta_k + 4 \tau \omega)}. \nonumber
\end{eqnarray}

In the low temperature limit ($\frac{T}{\varepsilon_F} \ll 1$), the oscillatory exponential varies fast but other terms change slowly. Therefore, we can treat only the oscillating exponential as a function of $k$ or $\varepsilon$. On the other hand, we keep only up to linear deviations for the expansion of the exponent near the Fermi energy. Then, Eq. (\ref{sigmaxxc0c2}) reads
\begin{eqnarray}
\sigma_{xx} &=& \sigma_{xx}^{l=0} -\frac{e^2k_F^2}{8 \hbar \pi^2} \sum_{l=1}^{\infty} M_l e^{\frac{\pi l i}{2\zeta_F}(1 + \zeta_F^2)} \int_{-\infty}^{\infty}\frac{e^{(1+i \lambda l/\pi)t}}{(e^t+1)^2}dt \nonumber \\
&=& \sigma_{xx}^{l=0} -\frac{e^2k_F^2}{8 \hbar \pi^2} \sum_{l=1}^{\infty} M_l \frac{\lambda l}{\sinh (\lambda l)}e^{\frac{\pi l i}{2\zeta_F}(1 + \zeta_F^2)} , \label{sigmaxxMl}
\end{eqnarray}
where $t \equiv \frac{\epsilon - \mu}{T}$, $\lambda \equiv \frac{\pi^2T}{\epsilon_F \zeta_F}(1-\zeta_F^2)$, and $M_l \equiv C_{0} - C_{2} \zeta_F e^{\frac{\pi i}{4}}/\pi l$.

Finally, inserting Eqs. (\ref{C0}) and (\ref{C2}) into Eq. (\ref{sigmaxxMl}), we find the transverse conductivity along the $x$ direction as
\begin{eqnarray}
\sigma_{xx} &=& \sigma_{xx}^{l=0} + 2 \sum_l \sigma_{xx}^{(l)}\left(\cos{\left(\frac{\pi l}{2\zeta_F}+\frac{\pi}{4}\right)}+\frac{l\pi}{\zeta_F}\cos{\left(\frac{\pi l}{2\zeta_F}-\frac{\pi}{4} \right)}\right), \nonumber \\
\end{eqnarray}
where
\begin{widetext}
\begin{eqnarray}
\sigma^{(l)}_{xx} &\equiv& \frac{n_e e^2 v_F}{\hbar k_F} \frac{(i\omega + \frac{3}{\tau})(1+\frac{1}{2}\frac{1+\omega \tau i}{1+(k_F v_F \tau \zeta_F)^2})+\frac{1}{\tau}\frac{(k_F v_F \tau \zeta_F)^2}{1+(k_F v_F \tau \zeta_F)^2}}{(i\omega + \frac{3}{\tau})^2 + (2 k_F v_F \zeta_F)^2} \frac{3}{2 \pi} \frac{\lambda l}{\sinh{\lambda l}}\left(\frac{2\zeta_F}{l}\right)^{\frac{3}{2}}\frac{1}{8} \nonumber \\
&\approx& \frac{n_e e^2 v_F}{\hbar k_F} \frac{1/\tau}{(i\omega + \frac{3}{\tau})^2 + (2 k_F v_F \zeta_F)^2} \frac{3}{2 \pi} \frac{\lambda l}{\sinh{\lambda l}}\left(\frac{2\zeta_F}{l}\right)^{\frac{3}{2}}\frac{1}{2}. \nonumber
\end{eqnarray}
\end{widetext}

\end{document}